\documentclass[12pt]{article}
\usepackage{amsmath, amsfonts,euscript,bbm}
\usepackage{epsfig, cite}
\usepackage{color}
\usepackage{ifthen}
\usepackage{graphicx}
\usepackage{setspace} 
\newboolean{Notes}\setboolean{Notes}{true}
\newcommand{\notes}[1]{\ifthenelse{\boolean{Notes}}{\textcolor{black}{#1}}{}}

\newcommand{\skyp}[1]{}

\def\lsim{\stackrel{<}{{_\sim}}}

\begin{document}

\bigskip
\hskip 4in\vbox{\baselineskip12pt \hbox{FERMILAB-PUB-08-353-A-T} }
\bigskip\bigskip\bigskip\bigskip\bigskip\bigskip\bigskip\bigskip

\centerline{\Large Spin-Statistics Violations in Superstring Theory}
\bigskip
\bigskip
\bigskip
\centerline{\bf Mark G. Jackson}
\medskip
\centerline{Particle Astrophysics Center and Theory Group}
\centerline{Fermi National Accelerator Laboratory}
\centerline{Batavia, Illinois 60510}
\medskip
\centerline{\it markj@fnal.gov}
\bigskip
\bigskip
\bigskip
\bigskip
\bigskip
\begin{abstract}
I describe how superstring theory may violate spin-statistics in an experimentally observable manner.  Reviewing the basics of superstring interactions and how to utilize these to produce a statistical phase, I then apply these ideas to two specific examples.  The first is the case of heterotic worldsheet linkings, whereby one small closed string momentarily enlarges sufficiently to pass over another, producing such a statistical phase.  The second is the braneworld model with noncommutative geometry, whereby matter composed of open strings may couple to a background in which spacetime coordinates do not commute, modifying the field (anti)commutator algebra.  I conclude with ways to sharpen and experimentally test these exciting avenues to possibly verify superstring theory.

\end{abstract}

\newpage            
\baselineskip=18pt
\section{Introduction}
A principle which has been very well-tested at low precision and energies is the Spin-Statistics Theorem (SST).  This states that given the assumptions of locality, Lorentz invariance and the vacuum being the lowest-energy state for a unitary point-particle field theory in 3+1 dimensions, integral-spin particles must be in a completely symmetric (`bosonic') wavefunctions whereas half-integral spin particles must be in completely antisymmetric (`fermionic') wavefunctions.  Despite many attempts at a simple proof (for an extensive review see \cite{Duck:1998cp}) there is none known, and so its validity is usually simply assumed when quantizing field modes.  This is done by imposing different (anti)commutators for the creation/annihilation operators:
\begin{equation}
\label{commut}
{\rm bosons:} \ [a_{\bf k},a^\dagger_{\bf p} ] = \delta_{{\bf k},{\bf p}}, \hspace{0.5in} {\rm fermions:} \   \{ b_{\bf k},b^\dagger_{\bf p} \} = \delta_{{\bf k},{\bf p}} .
\end{equation}
There are a variety of ways these relationships could be modified, as described in a review by Greenberg \cite{Greenberg:2000zy}, each of which requires relaxing at least one of the assumptions of locality, Lorentz invariance or the idea of a point-particle field theory altogether.  

Such violations of spin-statistics are theoretically interesting and could have dramatic physical and even cosmological consequences \cite{Jackson:2007tn}, but ideally they should be motivated by a UV-complete theory predicting such violations.  The leading such model, superstring theory, is fundamentally based upon extended objects and so clearly has the potential to produce such violations.  These could appear either at high energies or perhaps suppressed by some small but nonzero parameter in the theory.

In this article I will summarize two ways in which such violations might be produced in superstring theory.   $\S$2 contains a review of superstring interactions and how this produces a statistical phase.  In $\S$3 I present one specific way to possibly violate spin-statistics in the heterotic string theory through worldsheet linkings, and in $\S$4 a second method relying upon braneworlds and noncommutative geometry.  In $\S$5 I will offer some concluding remarks. 
\section{Superstring Interactions}
Both types of potential string theory spin-statistics violations come about from interactions with a gauge field, so we will briefly review this important issue here.
\begin{figure}
\begin{center}
\includegraphics[width=6in]{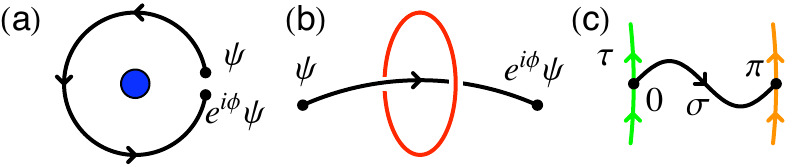}
\caption{(a) In 2+1 dimensions a charged particle's wavefunction will acquire a phase after a circuit around a flux tube, (b) A similar phase can be acquired in 3+1 dimensions for a particle passing through a closed loop of flux, or equivalently a string passing over a charge, or (c) by comparing the gauge field at the endpoints of an open string.}
\end{center}
\end{figure}
\subsection{Point Particles, Aharonov-Bohm, and Anyons}
Point particles naturally couple to the 1-form gauge field $A_\mu$ via an interaction on their worldline $X^\mu(l)$,
\[ S = q \int dl \ {\dot X}^\mu A_\mu. \]
This action is then combined with kinetic terms and inserted into a path integral over $X$ and $A$ to calculate scattering amplitudes,
\[ \mathcal A( \cdots) = \int \left[ \mathcal D A \right] \left[ \mathcal DX \right] e^{i S[A,X]} ( \cdots)  \]
where the $\cdots$ indicate field insertions.  Aharonov and Bohm first observed how such an interaction could be used to modify statistical phases in 2+1 dimensions \cite{Aharonov:1959fk}.  Consider two interacting particles, where the second one sources $A$ to produce some localized magnetic flux equal to $\Phi$, which in a particular gauge can be written
\[ A_i = - \frac{\Phi {\epsilon}_{ij} X^j}{4 \pi |X|^2}, \hspace{0.5in}  B_{12} = \Phi \delta^2(X) . \]
Now arrange for the first particle to perform a closed circuit around the second, as shown in Figure 1(a).  Such a circuit surrounding the source is topologically well-defined in the sense that one could smoothly adjust the path in an arbitrary fashion and yield the same enclosed flux.  Although the first particle is never in contact with the flux and so feels no force, it nonetheless induces a relative phase in the path integral
\[ \Delta \phi = q \int dl \ {\dot X}^i \left( - \frac{\Phi}{4 \pi} \epsilon_{ij} \partial^j \ln |X| \right) = q \Phi. \]
This phase will then modify the statistics (\ref{commut}), effectively violating spin-statistics and generalizing bosons and fermions into `anyons' \cite{Wilczek:1982wy}.  In order to evade the SST we had to break Lorentz invariance in the dimensional reduction.
\subsection{Superstrings and the Kalb-Ramond Field}
Similarly, superstrings naturally couple to the 2-form Kalb-Ramond gauge field $B_{\mu \nu}$ via the worldsheet interaction \cite{Rohm:1985jv} 
\begin{equation}
\label{stringaction}
S = \int d^2 z \ \partial X^\mu {\bar \partial X}^\nu B_{\mu \nu}.
\end{equation}
This is introduced into a path integral exactly the same as for a point particle and is also capable of producing a phase\footnote{While the action (\ref{stringaction}) is purely imaginary, the string path integral is Euclidean in the sense that it is introduced as $e^{-S}$ and so this action produces a phase just as for point particles.}.  

In the case of the string, however, there are now two possibilities, depending on whether the string is closed or open.  For a closed string, the topologically invariant quantity is not the amount of flux traversed in a particle's circular orbit but rather the amount of flux which has passed through the string loop \cite{Aneziris:1990gm} \cite{Bergeron:1994ym}, as shown in Figure 1(b); this is referred to as a `linking' in the literature.  One way this could happen is if a small string passed through a cosmic superstring \cite{Witten:1985fp} \cite{Copeland:2003bj} or other extended non-perturbative object \cite{Hartnoll:2006zb}.  While observing such an effect would be impressive, this isn't quite in the spirit of violating spin-statistics, and we would first have to find such a cosmically extended string! \cite{Polchinski:2004ia}  The second way is to begin with two small closed strings and allow one to momentarily enlarge; this is the mechanism elaborated upon in the next section.

Now specializing to open strings, such a linking is not possible, but there is a phase induced nonetheless.  In the case where $B$ is only defined at the string endpoints (as in $\S4$), the action and hence the phase will be equal to
\begin{equation}
\Delta \phi = \int d \tau \left. B_{\mu \nu}  {\dot X}^\mu {X}^{\prime \nu} \right|^{\pi}_{\sigma = 0} .
\end{equation}
This is shown in Figure 1(c).  Thus both types of string can produce statistical phases.  I will now apply these two cases to specific mechanisms leading to possible violation of spin-statistics.

\section{Method \#1: Heterotic Worldsheet Instantons}
\subsection{Motivation}
Let us consider two strings in 3+1 dimensions, where one is kept at finite size and the other is approximated as pointlike.   We have just seen that $B$-field flux passing through the string loop can produce a statistical phase, and so we desire that the second particle/string source this flux.  This can be achieved for a particle charged under a gauge field $A$ by using the (topological) interaction term of the form
\begin{equation}
\label{bf}
S_{BF} = \int d^4x \ \epsilon^{\mu \nu \rho \lambda} B_{\mu \nu} \partial_\rho A_\lambda
\end{equation}
which arises naturally from anomaly cancellation in the heterotic string \cite{Gross:1984dd}.  An instanton-like mechanism to utilize this fact was proposed by Harvey and Liu \cite{Harvey:1990wa}, whereby one string will momentarily open up and pass over another string before collapsing again, as shown in Figure 2.  The magnitude of this spin-statistics-violating effect was estimated to be of order $e^{-1/\alpha' E^2}$, assuming that one string must open up to at least the de Broglie wavelength of the other.  Naively $1/\sqrt{\alpha'} \sim 10^{16}$ GeV  and so this is prohibitively too small to be observed, but if $1/\sqrt{\alpha'} \sim$ TeV (as in some recent warped models \cite{Randall:1999ee} \cite{Kachru:2003sx} \cite{Curio:2000dw}) then perhaps this effect is observable at achievable energies and worth revisiting.  Note that this intrinsically stringy effect would never show up in the low-energy effective action, which is a Taylor expansion in small $\alpha'$.  
\begin{figure}
\begin{center}
\includegraphics[width=1.45in]{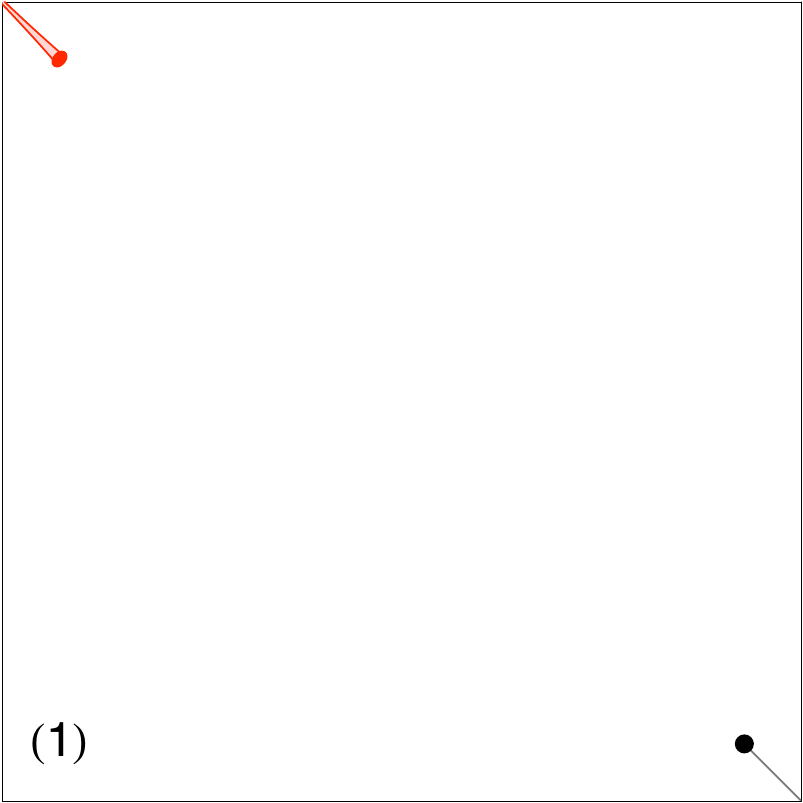}
\includegraphics[width=1.45in]{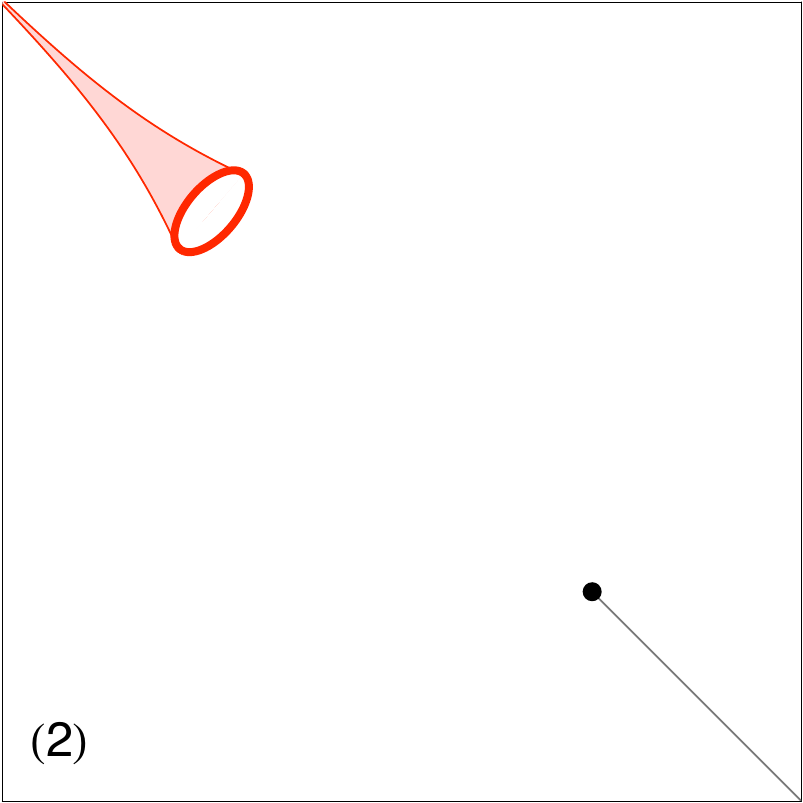}
\includegraphics[width=1.45in]{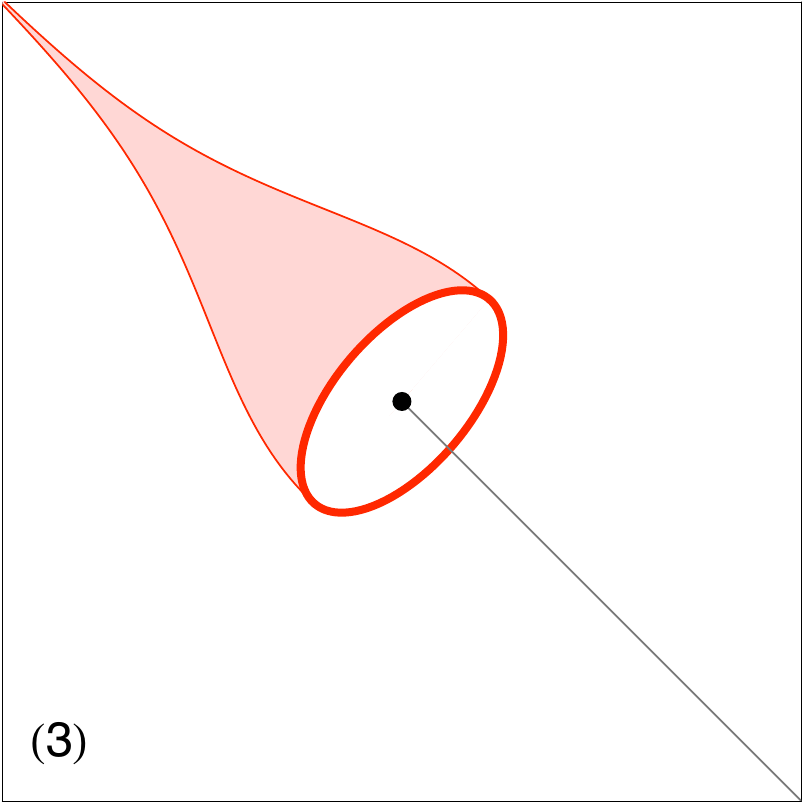}
\includegraphics[width=1.45in]{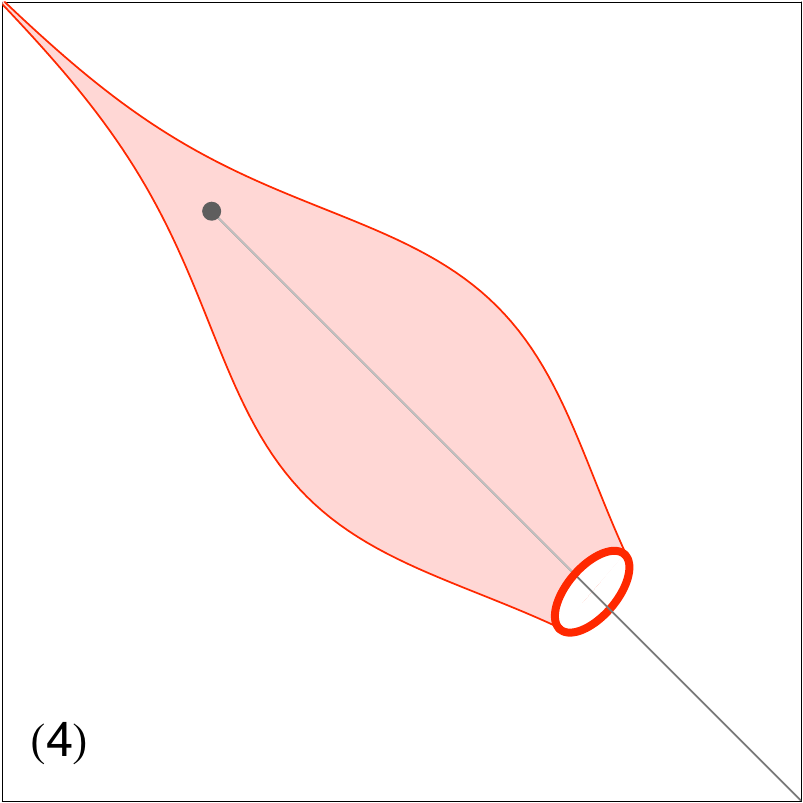}
\caption{Worldsheet instantonlike linking process whereby one string momentarily expands sufficiently to envelop another, producing a phase in the string path integral.}
\end{center}
\end{figure}

\subsection{Explicit Instanton Solutions}
Since the $BF$ term arises at 1-loop in a perturbative expansion in the string coupling, we would also expect the spin-statistics violation to occur at this order\footnote{I would like to thank S. Hellerman for discussions on this point.}.  While a detailed analysis is currently underway \cite{hellermanjackson}, one could try to estimate the magnitude of the effect by constructing instantonlike linking solutions as attempted in \cite{Jackson:2008bs}.  The complete action for the first string, with momentum $k_1$ and coupled to the Kalb-Ramond 2-form $B$, is
\[ S_{1} = \frac{1}{2 \pi \alpha'} \int d^2 z \left[ \partial X^\mu {\bar \partial} X^\nu (\delta_{\mu \nu} + 2 \pi \alpha' B_{\mu \nu}) + 2 \pi \alpha'  \delta^2(z, {\bar z}) k_1\cdot X  \right]. \]
Note again that the term containing $B$ is imaginary and thus produces a phase in the path integral, and that we are considering worldsheet instanton solutions so the momentum $k_1$ is real.  The action for the second string (which we approximate as a particle) with momentum $k_2$ coupled with charge $q$ to the pseudoanomalous $U(1)$ gauge field $A$ is
\[ S_2 = \int dl \ \left[  \frac{1}{2 \alpha'} {\dot Y} \cdot {\dot Y} + {\dot Y} \cdot \left( iq A-k_2 \right) \right]. \]
Again note that the term coupling to $A$ is imaginary.  The spacetime action governing the gauge fields $F=dA$ and ${\tilde H}=dB-A \wedge dA$ is the usual kinetic terms plus the $BF$ coupling in (\ref{bf}),
\[ S_{gauge} = \int d^4 x \left[ \frac{3 \alpha'}{32g^2} {\tilde H}^2 + \frac{1}{4g^2} F ^2 + \theta \epsilon^{\mu \nu \rho \lambda} B_{\mu \nu} \partial_\rho A_\lambda \right]   \]
where $g^2$ and $\theta$ are the dimensionless effective 4D couplings after compactification.   In the heterotic string theory with different compactifications we can get different values of $\theta = c/32 \pi^2$, where $c$ is determined by the massless fermion content of the theory.  In the case of compactification on a Calabi-Yau manifold \cite{Candelas:1985en} we break $SO(32) \rightarrow SU(3) \times SO(26) \times U(1)$ and then embed the spin connection in the gauge group.  This yields $c=-\frac{3}{2} \chi$, where $\chi$ is the Euler number of the Calabi-Yau, and the fermion charges are $q=\pm 1, \pm 2$.  The worldsheet and worldline then produce, respectively, $F$ and ${\tilde H}$ flux tubes with width $\sim \sqrt{\alpha'} /  \theta g^2$ (this is reversed from the usual case due to the $B F$ term).  If we approximate these as infinitesimally thin by taking $\theta g^2 \rightarrow \infty$ we may neglect the gauge kinetic terms and integrate the fields out, resulting in the effective action equal to
\begin{eqnarray*}
S_{eff} &=&  \frac{1}{2 \pi \alpha'} \int d^2 z\ | \partial (X - \alpha' k_1 \ln |z|)|^2 + \frac{1}{2 \alpha'} \int dl |\dot Y - \alpha' k_2|^2 + i \frac{qN}{\theta} \\
&=& \frac{1}{2 \pi \alpha'} \int d^2 z \ \left| \partial (X^\mu - \alpha' k_1^\mu \ln |z|) \right. \\
&& \left. \mp \ i \frac{ \pi qC\alpha' }{\theta} {\epsilon^\mu}_{ \nu \rho \lambda} \partial (X^\nu + \alpha' k_1^\nu \ln |z|) \int dY^\rho \partial^\lambda G(X-Y) \right|^2 \\
&+& \frac{1}{2 \alpha'} \int dl \ |\dot Y - \alpha' k_2|^2 + \frac{qN}{\theta} \left( i \pm C \right)
\end{eqnarray*}
where $N=\frac{\epsilon^{\mu \nu \rho  \lambda }}{4 \pi^2} \int d\Sigma_{\mu \nu}(X) \int dY_\rho \frac{ (X-Y)_\lambda}{|X-Y|^4}$ is the linking number.  The equation for $Y$ is trivial and yields $Y(l) = \alpha' k_2 l$, whereas that for $X$ is nontrivial and must first be transformed so that the derivative of $X$ is isolated on the LHS,
\begin{equation}
\label{bpsx}
z \partial X^\mu = \alpha'  \left( {\delta^\mu}_\nu + i \frac{ qC \alpha'}{4 \theta} {\epsilon^\mu}_{ \nu \rho \lambda}  \frac{ X_\perp^\rho {\hat k}_2^\lambda}{ |X_\perp |^3} \right)^{-1}  \left( {\delta^\nu}_\gamma - i \frac{ qC \alpha'}{4 \theta} {\epsilon^\nu}_{ \gamma \kappa \sigma}  \frac{ X_\perp^\kappa {\hat k}_2^\sigma}{ |X_\perp |^3} \right) k_1^\gamma.
\end{equation}
This can be shown to have no solutions except in the trivial case $X = \alpha' k_1 \ln |z|$, so that there exist no instanton solutions for the model above.

The reason for this is easy to understand: there is no force acting on the worldsheet to keep it open as it passes over the second string.  The most natural way to produce such an interaction is to recall that the (left-moving component of the) first string may also carry a charge $Q$ under the pseudoanomalous $U(1)$ gauge field, 
\begin{eqnarray*}
\label{ds1}
 \Delta S_1 &=&  \frac{1}{2 \pi} \int  d^2 z \ J(z) A_\mu {\bar \partial} X^\mu \\
 &\approx& i Q \int d \tau A_\mu {\dot X}^\mu 
 \end{eqnarray*}
where $J$ is the holomorphic $U(1)$ current normalized so that $\oint dz \ J(z) = 2 \pi i Q$.  Then the electrostatic repulsion between the two strings would expand the worldsheet to a radius 
\[ R \sim \sqrt{ g^2 q Q \alpha' } . \]
The addition of (\ref{ds1}) to the action for strings with $q Q > 0$ could then plausibly produce instanton solutions, and could be analyzed using techniques similar to those employed here.
Unfortunately explicit solutions for this model are likely much more difficult to construct due to the necessity of finite coupling.
\subsection{The Spacetime Effective Action}
Since explicit solutions may be difficult to obtain, let us for the moment assume that such solutions with linking number $N$ do exist and are of the Bogomol'nyi-Prasad-Sommerfeld form conjectured above, with a radially-symmetric trajectory producing this linking and an action proportion to $N$,
\begin{eqnarray}
\nonumber
X_N &=& \alpha' k_1 \ln |z| + f_N(|z|), \\
\label{sn}
S_N &=& \frac{q}{\theta} \left( i N+ C|N| \right) .
\end{eqnarray}
How would such a string theory process actually produce spin-statistics violations from the viewpoint of an effective field theory?  On one hand, such a violation is reasonable because spacetime spin-statistics only comes about in an indirect way in string theory, after Gliozzi-Scherk-Olive (GSO)-projection involving worldsheet spin-statistics (which are undisturbed even with spacetime background fields) \cite{GSW}.  Since the entire notion of vertex operators/GSO projection relies fundamentally on the fact that the string is an extended object, it is then reasonable to imagine that it could slightly violate spacetime spin-statistics.
On the other hand, string theory produces an effective action of Lorentz-invariant, local, point particle fields, which the SST demands obey usual spin-statistics.  How do we resolve this apparent paradox?

To see how, consider the correlation function between two strings in the background described by (\ref{sn}),
\begin{equation}
\label{a12}
 \mathcal A_{12} =  \int d^2z \sum_N \ e^{-k_2 \cdot \left[ \alpha' k_1 \ln |z| + f_N(|z|) \right] + iN/\theta - |N|C/\theta} 
 \end{equation}
where we have analytically continued back $X \rightarrow iX$ and summed over linking numbers.  To see what relation this has to the spacetime propagator, recall that the string propagator $\Delta$ can be represented in terms of worldsheet Hamiltonian $H = (p^2-m^2)$ and momentum $P$,
\[ \Delta = \frac{1}{2 \pi} \int_0 ^\infty d \tau \ e^{-H\tau} \int _{-\pi} ^\pi d \sigma \ e^{i \sigma P} . \]
Thus the correlation (\ref{a12}) represents states contracted via the effective propagator
\[ \Delta_{eff} = \frac{1}{2 \pi} \int_0 ^\infty d \tau \ e^{-H\tau} \sum_N e^{F_N(H,\tau) + iN/\theta - |N|C/\theta}  \int _{-\pi} ^\pi d \sigma \ e^{i \sigma P}  \]
where $z = e^{\tau + i \sigma}$ and $F_N$ is some function of both kinetic operators and worldsheet coordinates.  As $\theta \rightarrow 0$, only the $N=0$ term contributes and we recover the usual propagator.  Regardless of the details of the instantonlike solution, we see the propagator will necessarily be modified into something nonlocal, requiring an infinite number of derivatives.  It is probably not coincidence that such solutions likely require coupling to a gauge field via (\ref{ds1}), as happens in field theory \cite{Gulzari:2006sa}, allowing one to evade the spin-statistics theorem \cite{daCruz:2004si}:
\[ {\rm bosons:} \ \frac{1}{(p^2-m^2)^{1+\epsilon}}, \hspace{0.5in} {\rm fermions:} \ \frac{\displaystyle{\not} p+m}{(p^2-m^2)^{1+\epsilon}}, \hspace{0.5in} 0 < |\epsilon| \ll 1. \]
The introduction of nonlocal propagators in string theory has a precedent, but only on very unusual backgrounds \cite{Taylor:2003gn}.

It is important to stress that this nonlocality is not the usual nonlocality on size $\Delta x \sim \sqrt{\alpha '}$ because one has integrated out massive string modes.  This spin-statistics-violating nonlocality must be present at arbitrarily large distances, corresponding to adiabatically moving one particle around another\footnote{I am grateful to D. Tong for emphasizing this fact.}.
\subsection{Experimental Constraints}
It is difficult to place experimental constraints on such instanton effects without an explicit solution, since it is not clear whether the effect would scale non-perturbatively with energy or some small parameter such as coupling.  Therefore let us consider each:
\begin{itemize}
\item{\bf Energy scale} If the instantons scale with energy, as originally believed, it is possible that the LHC might see them if the effective string tension is very low, $\alpha' \sim (10 \ {\rm TeV})^{-2}$.  Also, some of the ``Transplankian" literature has discussed whether field-theory modifications like this could be observed in inflationary perturbations, such as Kempf's modified Heisenberg uncertainty \cite{Easther:2001fi}.   This could in principle probe (very) high energy, but there is not a great amount of precision data yet. 
\item{\bf Coupling constant} If the instantons instead scale with some small parameter such as a coupling constant, it is more likely that a precision experiment could see this.  Ramberg and Snow were the first to precisely measure possible deviations in fermionic spin-statistics \cite{Ramberg:1988iu}, at low energy but incredibly precise experiments detecting forbidden transitions.  Their approach has been refined by the VIP (VIolations of the Pauli exclusion principle) Experiment \cite{VIP} which has thus far constrained the deviation away from Fermi statistics in terms of the Ignatiev-Kuzmin-Greenberg-Mohapatra $\beta$ parameter \cite{Ignatiev:1987cd} \cite{Greenberg:1988um} as
\[ \frac{\beta^2}{2} \leq 4.5 \times 10^{-28}. \]
This bound is expected to improve another 2 orders of magnitude over the next few years due to larger integrated currents.  Though the energy scale is low at only 8 keV, the incredible precision means this might be a viable way of detecting  superstring-motived violations.
\end{itemize}

\section{Method \#2: Braneworlds and Noncommutative Geometry}
\subsection{Motivation}
The second such scenario we will study is that of brane worlds.  These are models in which our universe is represented as a D-brane whose worldvolume\footnote{We are ignoring the fact that in order for the matter to admit chiral representations of gauge symmetries, this D-brane must actually be the intersection of two higher-dimensional D-branes, but this is irrelevant for the present discussion.}  contains open strings representing Standard Model particles \cite{Blumenhagen:2005mu}, as shown in Figure 3.  The fact that the strings are open means that their boundary conditions are sensitive not just to the naive metric $g_{\mu \nu}$ but rather the metric and Kalb-Ramond $B$-field from before,
\[ \left. g_{\mu \nu} (\partial - {\bar \partial}) X^\nu + 2 \pi \alpha' B_{\mu \nu} ( \partial + {\bar \partial}) X^\nu \right|_{z={\bar z}} = 0. \]
This generally difficult set of boundary conditions can be simplified by identifying $\theta^{\mu \nu} = \left( B^{-1} \right)^{\mu \nu}$ while simultaneously taking the string tension $\alpha' \sim \sqrt{\epsilon} \rightarrow 0$ and the metric $g_{\mu \nu} \sim \epsilon \rightarrow 0$ \cite{Seiberg:1999vs}.  This limit imposes noncommutative geometry in the sense that fields corresponding to these open strings are now multiplied by the Moyal star product $\star$ defined as 
\begin{equation}
\label{moyaldef}
\star \equiv e^{- \frac{i}{2} \theta^{\mu \nu} P_\mu  P_\nu }
\end{equation}
which for spatially-dependent fields means it acts as
\[ \phi(x) \star \Phi(y) =  e^{\frac{i}{2} \theta^{\mu \nu} \frac{\partial}{\partial x^\mu}  \frac{\partial}{\partial y^\nu} } \phi(x) \Phi(y) . \]
This algebra can be summarized by stating that spatial coordinates fail to commute by a constant $\theta$,
\[ [{\hat x}^\mu, {\hat x}^\nu] = i \theta^{\mu \nu}. \]
Since the noncommutative coordinates now produce nonlocal interactions it is reasonable that this might violate spin-statistics, especially given the close connection of noncommutative geometry and the Quantum Hall Effect \cite{Susskind:2001fb} which relies fundamentally on such violations.
\begin{figure}
\begin{center}
\includegraphics[width=3in]{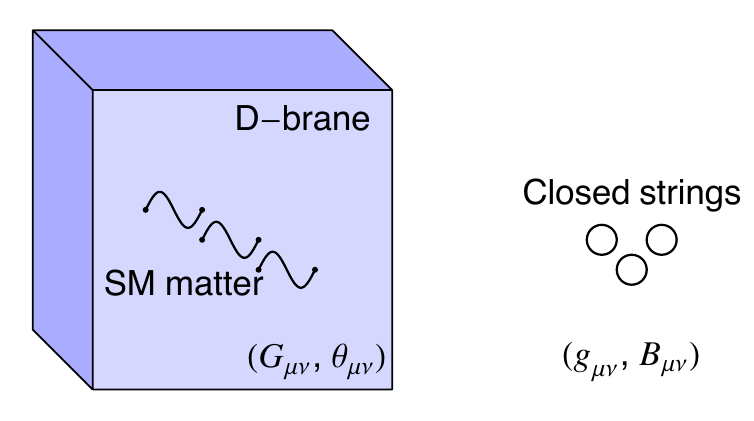}
\caption{The Braneworld model considers our universe to be a D-brane, and Standard Model particles are open strings whose endpoints are stuck on the brane.  Closed strings outside the brane see the fields $g_{\mu \nu}$ and $B_{\mu \nu}$, but open strings on the brane instead see a different metric $G_{\mu \nu}$ and noncommutativity parameter $\theta_{\mu \nu}$.}
\end{center}
\end{figure}
\subsection{Spin-Statistics Violations from Noncommutativity}
This idea has been studied in a series of a papers \cite{Chaichian:2002vw} \cite{Balachandran:2005eb} \cite{Tureanu:2006pb} which we will now summarize.  To see that noncommutative geometry is non-local, consider the equal-time commutator matrix element between the vacuum and a 2-particle state in $d$-dimensions, which should vanish in a local theory since measurements outside the lightcone can't influence each other:
\begin{eqnarray*} 
&& \langle 0 | \left. [ : \phi(x) \star \phi(x) :, : \phi(y) \star \phi(y) : ] \right|_{x^0 = y^0} | p,p' \rangle \\
&=& - \frac{2i}{(2 \pi)^{2d}} \frac{1}{ \sqrt{\omega_p \omega_{p'} }} \left( e^{-i p'x - i py} + e^{-ipx - ip'y} \right)  \int \frac{ d^3 k}{\omega_k} \sin \left[ k \cdot (x-y) \right] \cos \left( \frac{1}{2} k \cdot \theta \cdot p \right) \cos \left( \frac{1}{2} k \cdot \theta \cdot p' \right) .
\end{eqnarray*}
If $\theta^{i0}=0$ then the integrand is antisymmetric under $k^i \rightarrow -k^i$, and so the correlation vanishes upon integration over the spatial measure $d^3 k$.  Thus only if $\theta^{i 0} \neq 0$ could such a nonlocal process occur.  Unfortunately, such timelike noncommutative theories are found to violate unitarity, the reason being that they cannot be formulated as field theories decoupled from massive open string modes \cite{Gomis:2000zz}.  This does not mean all hope is lost, however.  There is a technical loophole in that \emph{lightlike} noncommutativity, for which $\theta^{\mu \nu} \theta_{\mu \nu} = 0$, produces theories which are indeed unitarity! \cite{Aharony:2000gz}  There has been relatively little work done in this but hopefully the motivation from spin-statistics violations will spark some interest.

One might instinctively guess that the mixing of the coordinates would produce spin-statistics violations, and so this may come as a rather counterintuitive conclusion.  Let us verify the matter explicitly for only $\theta^{ij} \neq 0$.  Consider a real scalar field $\phi$,
\begin{equation}
\label{phi}
 \phi(x) = \int \frac{d^3 k}{(2 \pi)^{3/2}} \left( a_{\bf k} e^{-i {\bf k} \cdot {\bf x}} + a^\dagger_{\bf k} e^{i {\bf k} \cdot {\bf x}} \right). 
\end{equation}

For the Moyal $\star$-product we can again choose the spatial representation $P_i \rightarrow - i \partial_i$, in which case it is simply the Fourier phases which are multiplied:
\begin{eqnarray}
\nonumber
\phi(x) \star \phi(y) &=& \int d^3 k \ d^3 p \ {\tilde \phi}(k) {\tilde \phi} (p) \left( e^{-ikx} \star e^{-ipy} \right) \\
\label{moyal}
&=& \int d^3 k \ d^3 p \ {\tilde \phi}(k) {\tilde \phi}(p) e^{-ikx - ipy + {1 \over 2} k \theta p}
\end{eqnarray}
and the raising and lowering operators will still obey the usual algebra $[a_{\bf k}, a^\dagger_{\bf p}] = \delta_{{\bf k},{\bf p}}$.  So from this perspective we get a noncommutative theory but one which respects the usual spin-statistics relation.

In \cite{Balachandran:2005eb} it is claimed that one could make an alternate choice of the Moyal Star representation, since the Fourier modes ${\tilde \phi} (k)$ of a field $\phi(x)$ also furnish representations of the momentum generators $P^i$:
\[ P^i {\tilde \phi} (k) = k^i  {\tilde \phi} (k) . \]
Given the mode expansion (\ref{phi}) this can be interpreted as deformed operators $a_{\bf k}, a^\dagger_{\bf k}$ relative to the undeformed ones $c_{\bf k}, c^\dagger_{\bf k}$,
\[ a_{\bf k} = c_{\bf k} e^{- \frac{i}{2} p_\mu \theta^{\mu \nu} P_\nu }, \hspace{0.5in} a^\dagger_{\bf k} = e^{\frac{i}{2} p_\mu \theta^{\mu \nu} P_\nu } c^\dagger_{\bf k} \]
which will then produce the following deformed commutation relations
\begin{eqnarray*}
a_{\bf k} a_{\bf p} &=& e^{-i p \cdot \theta \cdot k} a_{\bf p} a_{\bf k}, \hspace{0.5in} a_{\bf k}^\dagger a_{\bf p}^\dagger = e^{-i p \cdot \theta \cdot k} a_{\bf p}^\dagger a_{\bf k}^\dagger, \\
a_{\bf k} a_{\bf p}^\dagger &=& e^{i p \cdot \theta \cdot k} a_{\bf p}^\dagger a_{\bf k} + 2 E_{\bf k} \delta^3 ({\bf p}-{\bf k}).
\end{eqnarray*}
A field quantized with these deformed commutation relations will undo the $\star$ operation in (\ref{moyal}), and thus render the $S$-matrix identical to that for a standard (commuting geometry) field, suggesting that a noncommutative theory with usual spin-statistics could be interpreted as a commuting theory with modified spin-statistics.  In fact this is not true, as detailed in \cite{Tureanu:2006pb}.  Were we to include the Fourier components in the noncommutative field multiplication, there must now be \emph{three} Moyal star multiplications required:
\[ \phi(x) \star \phi(y) = \int d^3 k \ d^3 p \ {\tilde \phi}(k) \star e^{-ikx} \star {\tilde \phi} (p) \star e^{-ipy} . \]
The first and third $\star$ operations are trivial, but the second will produce the identical result as that obtained in (\ref{moyal}).  Thus the theory is truly noncommutative, and obeys the standard spin-statistics relations.
\subsection{Experimental Constraints}
For this model violations of spin-statistics are parameterized in terms of the noncommutativity parameter $\theta$.  Besides the usual constraints on Lorentz violation\cite{Kostelecky:2002hh} \cite{Kostelecky:2003fs}, there exist bounds on the spatial components as $|\theta^{XY}| \lsim (10^{14} \ {\rm GeV})^{-2}$ from QCD \cite{Mocioiu:2000ip} and $|\theta^{XY}| \lsim (10 \ {\rm TeV})^{-2}$ from QED \cite{Carroll:2001ws}.  Constant $H = dB$ has also been studied \cite{Nastase:2006na}, finding that it would behave as stiff matter $\rho_H \sim a^{-6}$ but with strange properties such as solitons moving arbitrarily faster than the speed of light, and so the amount of such flux must be limited.  Finally, there has been so-called Transplanckian research specifically studying whether noncommutative geometry might be measurable in the cosmic microwave background power spectrum \cite{Chu:2000ww}.

These constraints on the spatial components should provide some constraints on the lightlike components\footnote{I would like to thank A. Kostelecky for private communication on this point.}.  However, it was noted in \cite{Aharony:2000gz} that for lightlike $\theta^{i-}$ there is apparently no meaningful way to parameterize such violation, since the Minkowski square of such a quantity will always be zero by definition!  Thus it is difficult to say whether such a background exists, making the theoretical and experimental method to constrain lightlike noncommutativity an important challenge.  

\section{Conclusion, Future Directions and Open Questions}
There are several interesting theoretical and experimental issues that need to be addressed in the context of superstring violations of spin-statistics:
\begin{enumerate}
\item Does quantum gravity manifest itself as violations of spin-statistics?  If so, is there a simple way to encode this new physics? Greenberg \cite{Greenberg:2000zy} makes the interesting observation that such violations cannot be encoded into an effective action as a ``statistics violating term," so perhaps this is one reason we have found it difficult to quantize gravity?
\item Is there a mechanism in string theory to produce such violations, and is it related to the Kalb-Ramond field $B$? Both mechanisms mentioned here directly involve the $B$ field because this is the simplest way to introduce a phase, but are there others?  Would a greater understanding of why we don't observe such $B$-quanta help?  Note also that after compactification to 4 dimensions this field is actually an axion (since $d a = \ast d B$), so perhaps this is related to axion physics \cite{Svrcek:2006yi}.
\item Are such violations way beyond any possible experiment, or are they within reach of current technology? (such as precise but low-energy experiments like VIP or less-precise but higher-energy experiments like the LHC).  Do the violations scale with energy, fixed parameters, or a lightlike noncommutative parameter?  If the latter, how would we parameterize it?
\end{enumerate}
I believe that attempting to answer these questions will prove to be an important step in understanding the relationship between quantum gravity and possibly proving superstring theory.
\section{Acknowledgments}
I would like to thank J. Gomis, S. Hartnoll, J. Harvey, S. Hellerman, C. Hogan, A. Kostelecky, J. Lykken, C. Petrascu and D. Tong for useful discussions and collaborations.  This work was supported by the DOE at Fermilab.

\end{document}